\documentclass[intlimits,twoside,a4paper]{article}

\usepackage{amsmath,amssymb}
\usepackage{graphicx}
\usepackage{caption}
\usepackage{subcaption}

\usepackage[T2A]{fontenc}
\usepackage[cp1251]{inputenc}

\usepackage[eqsecnum]{cmpj2}

\issue{2015}{18}{1}{13004}
\doinumber{10.5488/CMP.18.13004}

\title[Effects of translational and rotational freedom]%
{Effects of translational and rotational degrees of freedom on the properties of model water}

\author[T. Mohori\v{c}, B. Hribar-Lee, V. Vlachy]{T. Mohori\v{c}, B. Hribar-Lee, V. Vlachy}
\address{Faculty of Chemistry and Chemical Technology, University of Ljubljana, \\ Ve\v{c}na pot 113, SI-1000 Ljubljana, Slovenia}

\date{Received October 24, 2014, in final form December 18, 2014}

\begin{document}

\maketitle

\begin{abstract}
Molecular dynamics simulations with separate thermostats for
rotational and translational motions were used to study the effects
of these degrees of freedom on the structure of water at a fixed
density. To describe water molecules, we used the SPC/E model.
The results indicate that an increase of the rotational
temperature, $T_\textrm{R}$, causes a significant breaking of the
hydrogen bonds. This is not the case, at least not to such
an extent, when the translational temperature, $T_\textrm{T}$, is raised.
The probability of finding an empty spherical cavity (no water
molecule present) of a given size, strongly decreases with an
increase of $T_\textrm{R}$, but this only marginally affects the free
energy of the hydrophobe insertion. The excess internal energy
increases proportionally with an increase of $T_\textrm{R}$, while an
increase of $T_\textrm{T}$ yields  a much smaller
effect at high temperatures. The diffusion coefficient of water exhibits a
non-monotonous behaviour with an increase of the rotational
temperature.

\keywords water structure, degrees of freedom,
molecular dynamics

\pacs 05.20.Jj, 05.70.Ln, 61.20.Ja, 61.20.Qg, 61.25.Em
\end{abstract}

\section{Introduction}

We dedicate this contribution to J.A. Barker and D. Henderson
\cite{Barker}, authors of the important paper {\it What is
``liquid''? Understanding the states of matter}, which was
published in Reviews of Modern Physics, almost exactly 50 years
ago. This excellent paper, which served as a textbook to many
of us, is still a reference article for every person who wishes
to study the theory of liquids. Our contribution here touches
upon one particular, but extremely important liquid~--- water,
from the standpoint of the individual degrees of freedom.

Complex molecules have, besides the translational, also
rotational and vibrational degrees of freedom. They all make
their own contribution to thermodynamic functions, which depend
on temperature of the system. Temperature is accordingly
considered to be the most important thermodynamic parameter. An
interesting question is how the fundamental degrees of freedom
contribute to the potential of mean force between water
molecules, when their characteristic temperatures vary
independently.

A biologically important situation arises when water molecules
are irradiated with microwaves of appropriate frequency to
excite their rotational motion, which affects the solvation and
interaction between the solute molecules
\cite{Bren2010,Bren2012,Mohoric2014}. The interaction of
microwaves with water has been previously studied using the
non-equilibrium molecular dynamics simulation
\cite{Tanaka2007}, where the electric field was modelled
explicitly. It has been shown that microwaves induce heating by
excitation of rotational motion in water molecules. The excess
rotational kinetic energy is then transferred to translational
degrees of freedom. Following previous studies
\cite{Bren2010,Bren2012,Mohoric2014}, the direct interaction of
microwaves with a solvent is replaced by a model system in which
the rotational degree of freedom has a  different
temperature than the translational motion. Such a
non-equilibrium steady-state picture, where different degrees
of freedom can have different temperatures, has already been
used to explain experimental phenomena, such as increased
chemical reaction rates and altered polar solvent boiling
points \cite{Bren2008,Bren2010}.

In the preceding contribution \cite{Mohoric2014}, we studied the
effects in hydration of cations, anions and hydrophobes caused
by independent variations of the rotational and translational
temperatures. The main conclusions were: (i) an increase of the
rotational temperature affects the hydration of cations in an
opposite way to anions, (ii) an increase of the translational
temperature always decreases the height of the first peak in
the solute--water radial distribution function; (iii) an
increase of the rotational temperature yields an increase in
the first peak in the solute--water radial distribution function
for hydrophobes and cations; while, (iv) in contrast to this,
the solvation peak decreases around ions with a sufficiently
large negative charge.

The focus of the preceding work was, therefore, on solvation of
various solutes and not much attention was paid to the solvent
itself.  With this respect, the present work complements the
previous one; here, we are primarily interested in the structure of
the water under such non-equilibrium conditions. In order to
explore the properties of the model water, we used molecular
dynamics (MD) simulations, where the rotational and
translational motions were coupled to separate thermostats. We
used a rigid model of water (SPC/E) \cite{Berendsen1987}; the
effect of vibrational temperature variations was found
to be small \cite{Bren2012}. The focus of the present study is
on the structure of water as reflected in various site--site
distribution functions. The probability of water molecule to
form a certain number of hydrogen bonds is calculated as a
function of rotational and translational temperatures. The
effects of the separate $T_\textrm{R}$ and $T_\textrm{T}$  variations on the
excess internal energy of the system and on the dynamics of water
molecules is examined.

\section{Model and simulations}

Molecular dynamics simulations were performed at a constant
number of particles and volume ($N,V$) with 256 SPC/E water
molecules in the simulation box. As before \cite{Mohoric2014},
the number density of water mole\-cules was set to 1.0~g/cm$^3$,
and the time step used was 1.0~fs. Each MD simulation had at
least 100~ps long MD equilibration run, while the statistics
was collected during the 4900~ps long production run.

The translational motion of a single water molecule $i$ was
described by its center-of-mass position $\textbf{r}_i$,
velocity $\textbf{v}_i$ and the force $\textbf{f}_i$ acting on
it. Similarly, the rotational motion of a single water molecule
$i$ was characterized by its orientation (using quaternion
representation) $\textbf{q}_i$, angular velocity
$\boldsymbol\omega_i$ and torque $\boldsymbol\tau_i$. While the
acceleration of the center-of-mass follows immediately from the
force acting on it, the angular acceleration must be computed
from Euler's equations of motion for rigid bodies. We employed
the standard Verlet algorithm to integrate the equations of
motion \cite{Allen1987}. To hold $T_\textrm{R}$ and $T_\textrm{T}$ fixed at their
respective values, we employed a simple velocity re-scaling
scheme.

\section{Results and discussion}

\subsection{Radial distribution functions}

\begin{figure}[!t]
\centering
\begin{subfigure}[b]{0.5\textwidth}
	\centering
	\includegraphics[keepaspectratio=true,angle=270,width=\textwidth]{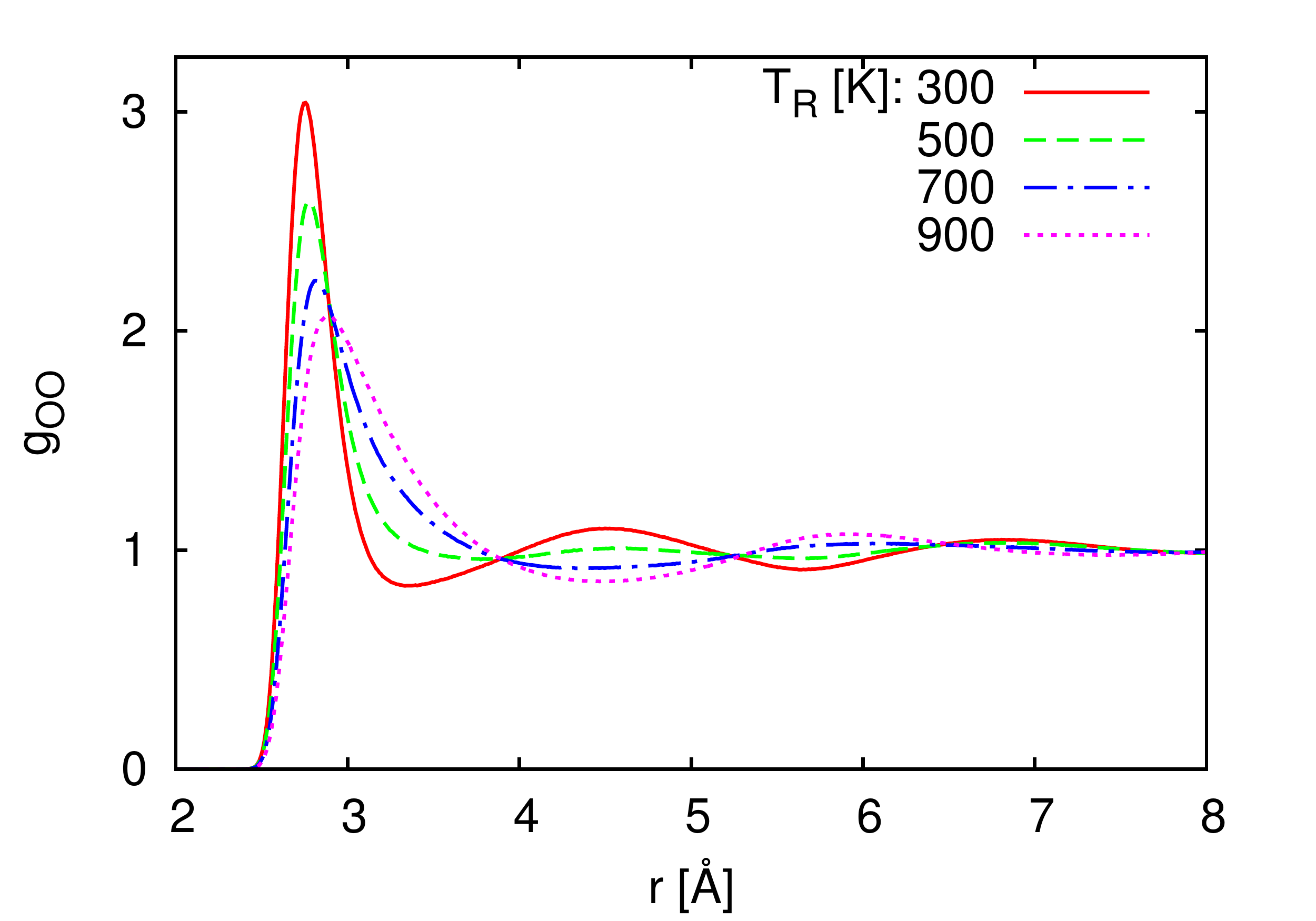}
\end{subfigure}%
\begin{subfigure}[b]{0.5\textwidth}
	\centering
	\includegraphics[keepaspectratio=true,angle=270,width=\textwidth]{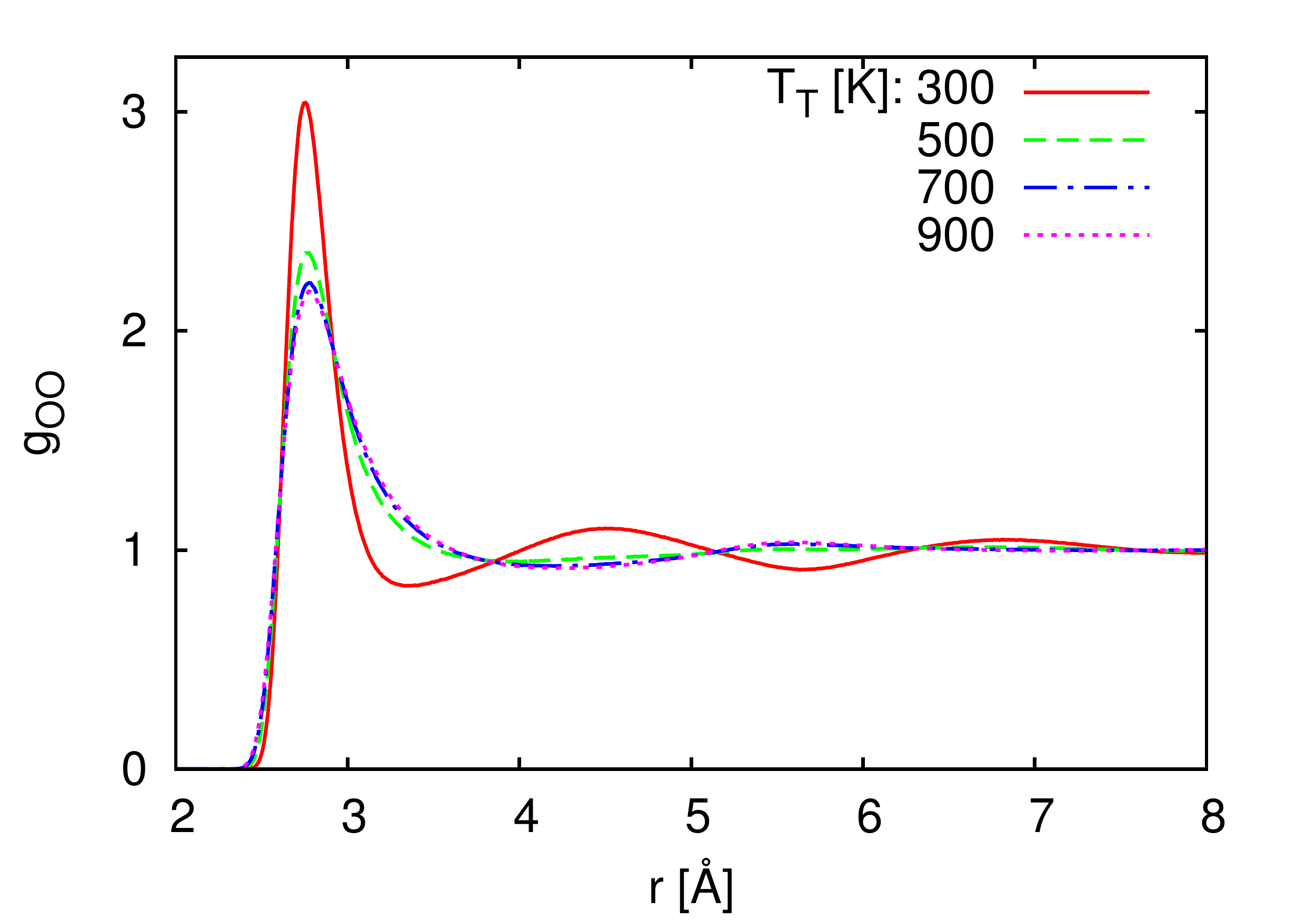}
\end{subfigure}
\caption{(Color online) Oxygen--oxygen radial distribution functions ($g_\textrm{OO}$) for different $T_\textrm{R}$ (left-hand) and $T_\textrm{T}$ (right-hand) values. Different lines correspond to different
values of the relevant temperature: 300~K (solid, red), 500~K (dashed, green), 700~K (dash-dotted blue), and 900~K (dotted, magenta).}
\label{pdf-1}
\end{figure}

We start the presentation of numerical results with the
oxygen--oxygen, $g_\textrm{OO}$, and oxygen--hydrogen distribution
functions, $g_\textrm{OH}$. First, in figure~\ref{pdf-1} we
present the $g_\textrm{OO}$ distributions for various rotational
($T_\textrm{R}$) and translational ($T_\textrm{T}$) temperatures of the system.

In both cases, an increase of the temperature decreases the
height of the first peak of the oxygen--oxygen distribution
function. However, there is a difference regarding how the position of
the peak responds to the temperature variations. An increase of
$T_\textrm{R}$ causes the first peak to move toward larger
distances. In contrast to this, the position of the first peak
in $g_\textrm{OO}$ does not change upon an increase of the
translational temperature, $T_\textrm{T}$. In addition, the peak becomes
somewhat broader; the shape seems to reflect the fact that
molecules with larger kinetic energy can easier ``penetrate''
into each other.

\begin{figure}[!t]
\centering
\begin{subfigure}[b]{0.5\textwidth}
	\centering
	\includegraphics[keepaspectratio=true,angle=270,width=\textwidth]{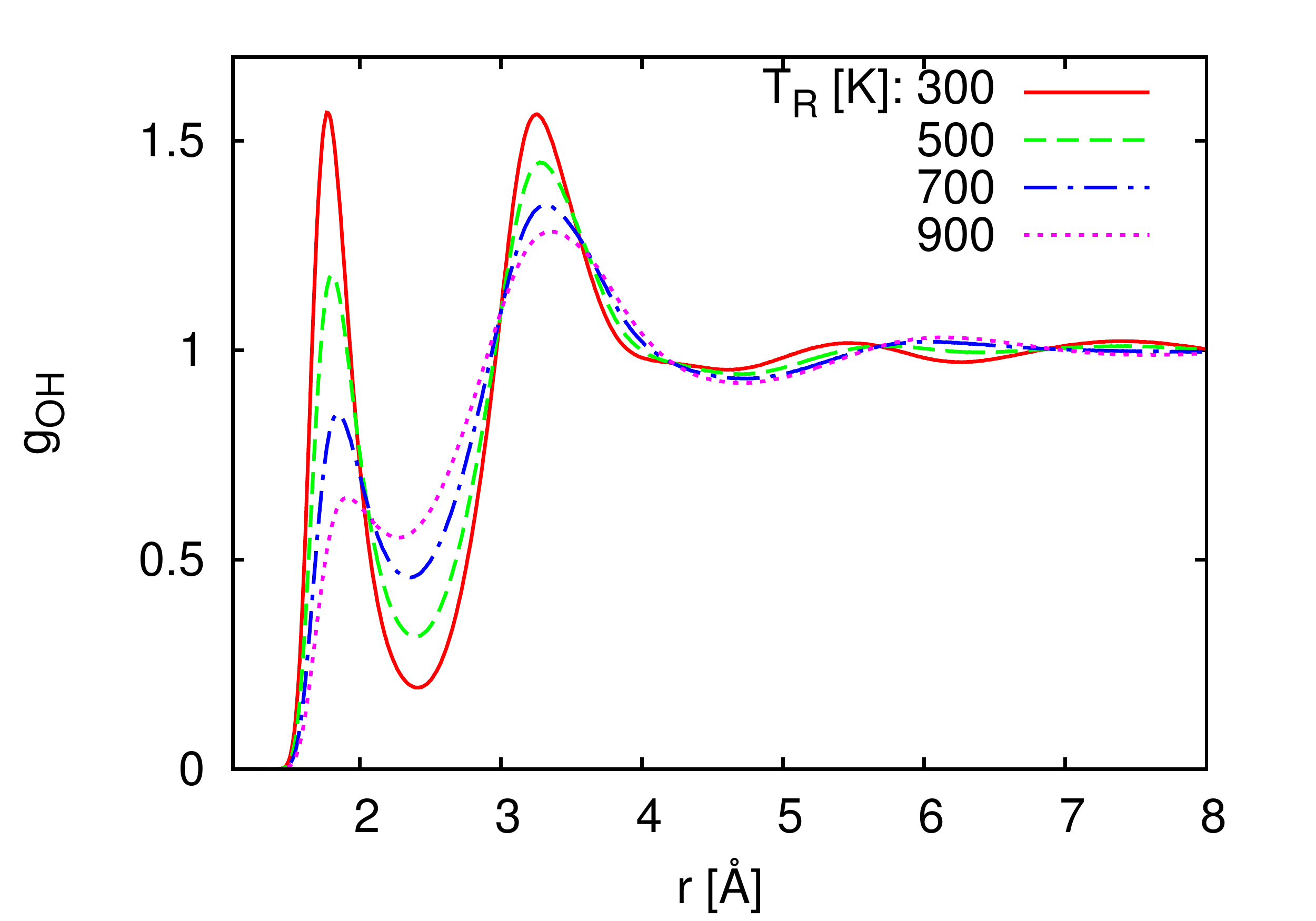}
\end{subfigure}%
\begin{subfigure}[b]{0.5\textwidth}
	\centering
	\includegraphics[keepaspectratio=true,angle=270,width=\textwidth]{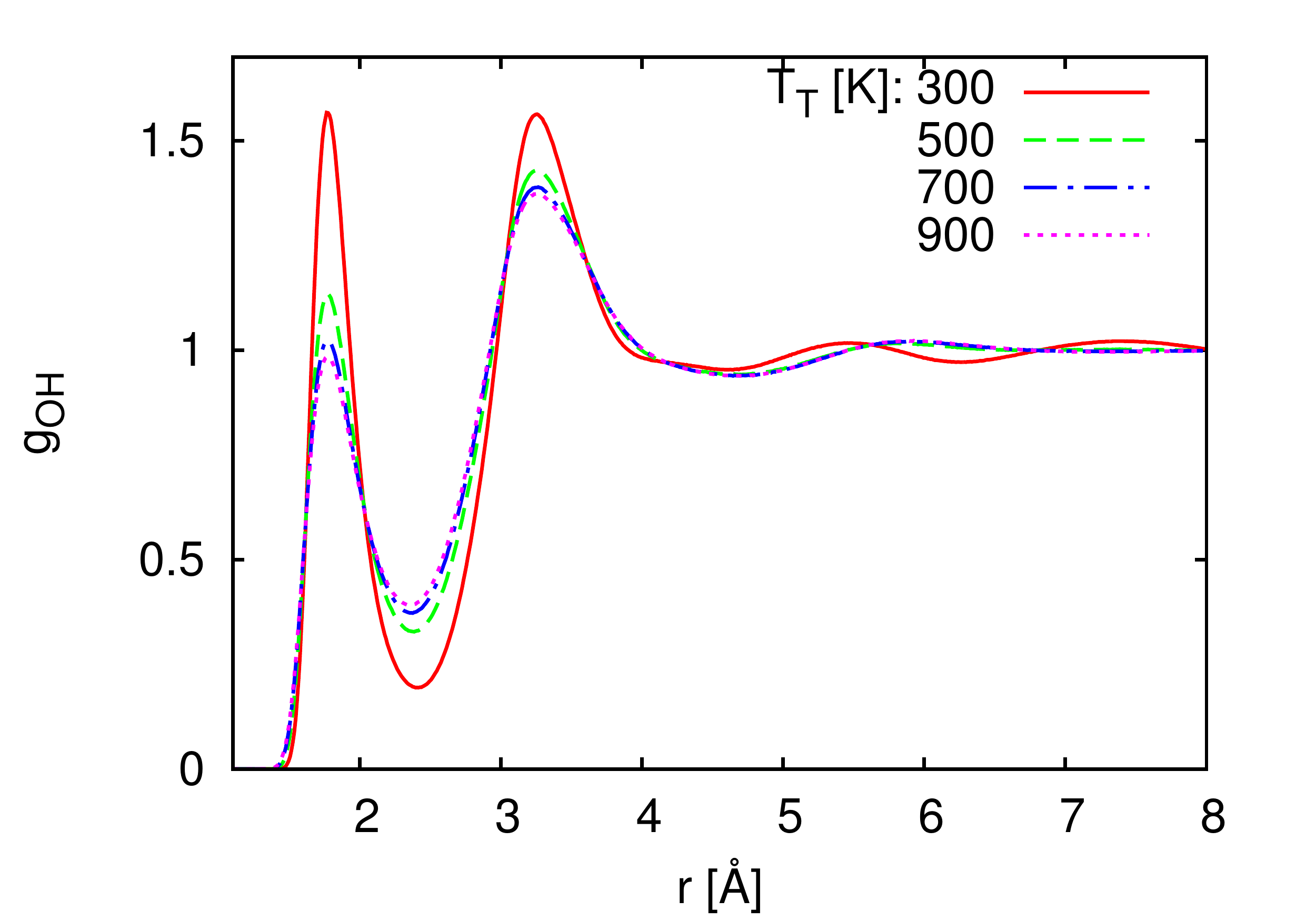}
\end{subfigure}
\caption{(Color online) Oxygen--hydrogen radial distribution functions ($g_\textrm{OH}$) for different $T_\textrm{R}$ (left-hand) and $T_\textrm{T}$ (right-right) values. The color code as for figure~\ref{pdf-1}.}
\label{fig-2}
\end{figure}

\begin{figure}[!h]
\centering
\includegraphics[keepaspectratio=true,angle=270,scale=0.3]{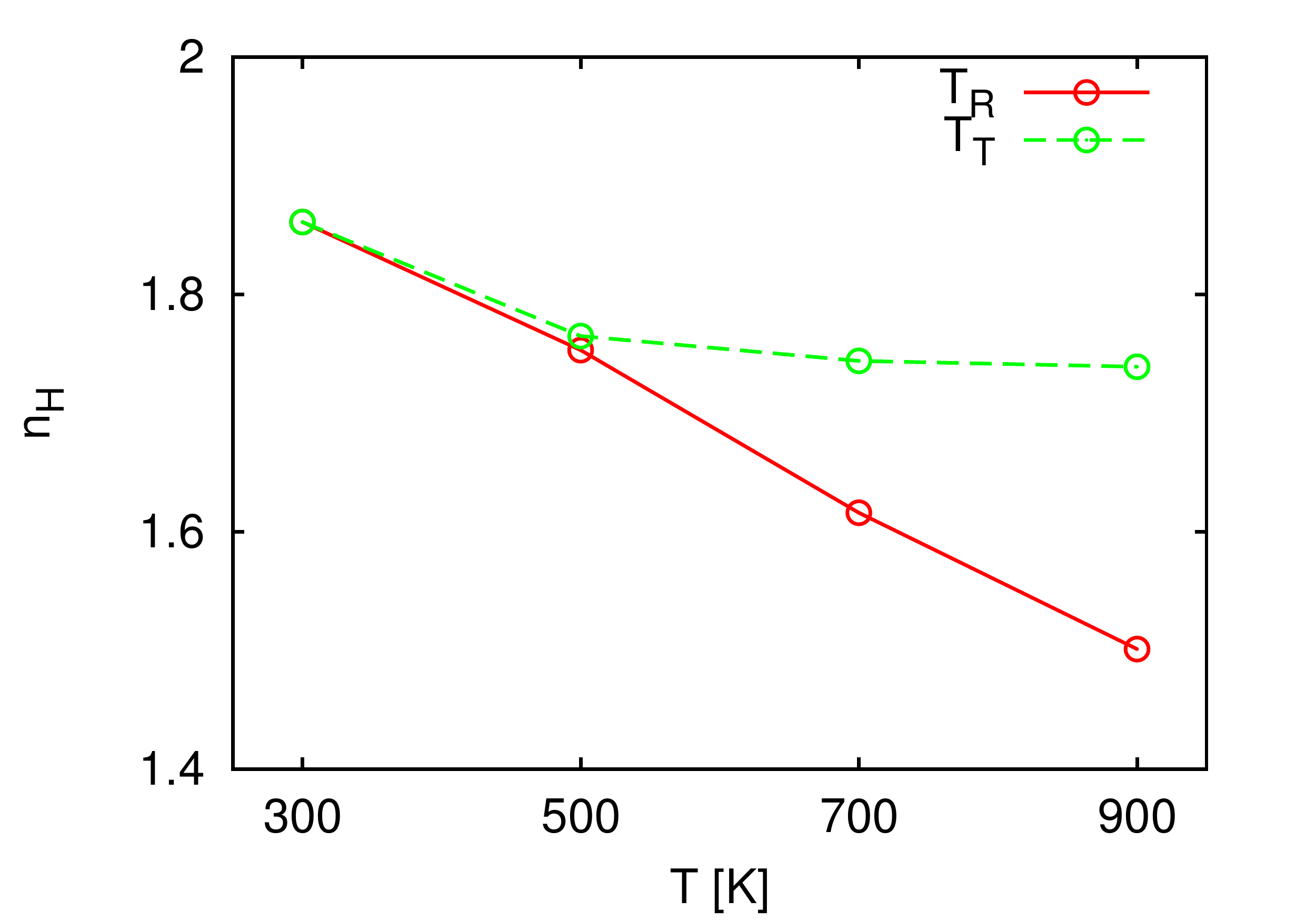}
\caption{(Color online) Coordination number, $n_\textrm{H}$, of hydrogen atoms of the neighbouring molecules around oxygen of the central molecule (and {\it vice versa}) as a function of $T_\textrm{R}$ (solid, red) and $T_\textrm{T}$ (dashed, green). $n_\textrm{H}$ is obtained as the volume integral of $g_\textrm{OH}$ up to the first minimum (at 2.4~{\AA}).}
\label{coord_num}
\end{figure}

Liquid water is at ambient conditions characterized by the
position of the second peak in this figure (figure~\ref{pdf-1}). The peak is located at 4.5~{\AA} rather than at
$2\sigma_{WW}\sim 6$~{\AA} as it is at the absence of hydrogen bonds.
The position of this peak is a consequence of the
hydrogen-bonding of water molecules. The shift of the second peak
toward $ \sim 6$~{\AA}  suggests that some hydrogen bonds
break upon an increase of rotational temperature.
The effects of temperature variations on the hydrogen bonds in
the model water are most clearly shown in figure~\ref{fig-2}. Here,
the oxygen--hydrogen distribution, $g_\textrm{OH}$, is presented as a
function of the respective temperature. The position of the
first peak of $g_\textrm{OH}$ distribution is characteristic of
a hydrogen bond. We see that the height of this peak is very
sensitive to $T_\textrm{R}$ and much less to the $T_\textrm{T}$ variations. In
both cases, however, the peak decreases in magnitude. From this
distribution, the coordination number, $n_\textrm{H}$, is calculated
(for the definition see caption in figure~\ref{coord_num}).
This quantity, as shown in figure~\ref{coord_num},
decreases almost linearly with $T_\textrm{R}$. On the other hand, $n_\textrm{H}$
is much less sensitive to the $T_\textrm{T}$
variations at higher temperatures.

\begin{figure}[!t]
\centering
\begin{subfigure}[b]{0.5\textwidth}
	\centering
	\includegraphics[keepaspectratio=true,angle=270,width=\textwidth]{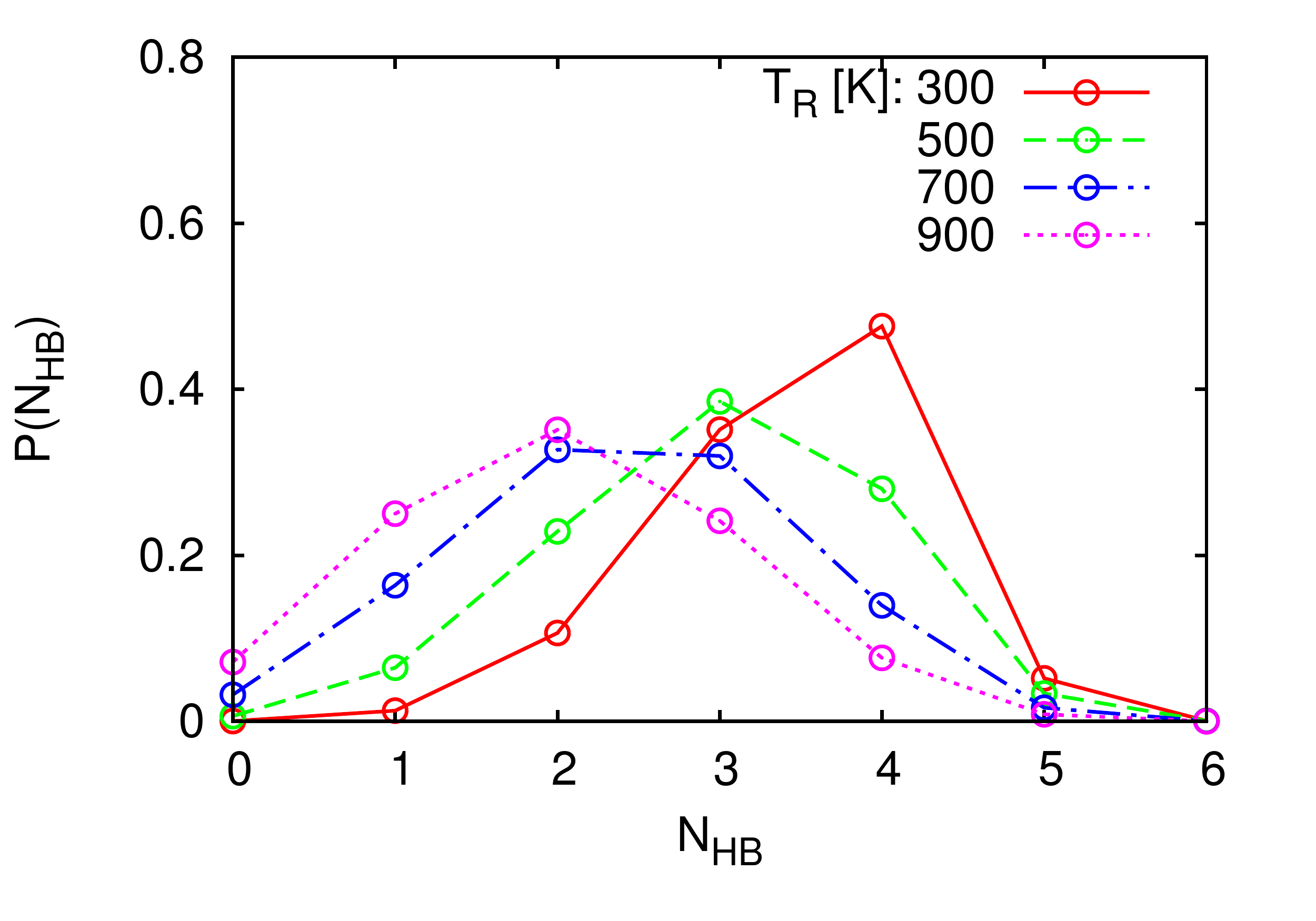}
\end{subfigure}%
\begin{subfigure}[b]{0.5\textwidth}
	\centering
	\includegraphics[keepaspectratio=true,angle=270,width=\textwidth]{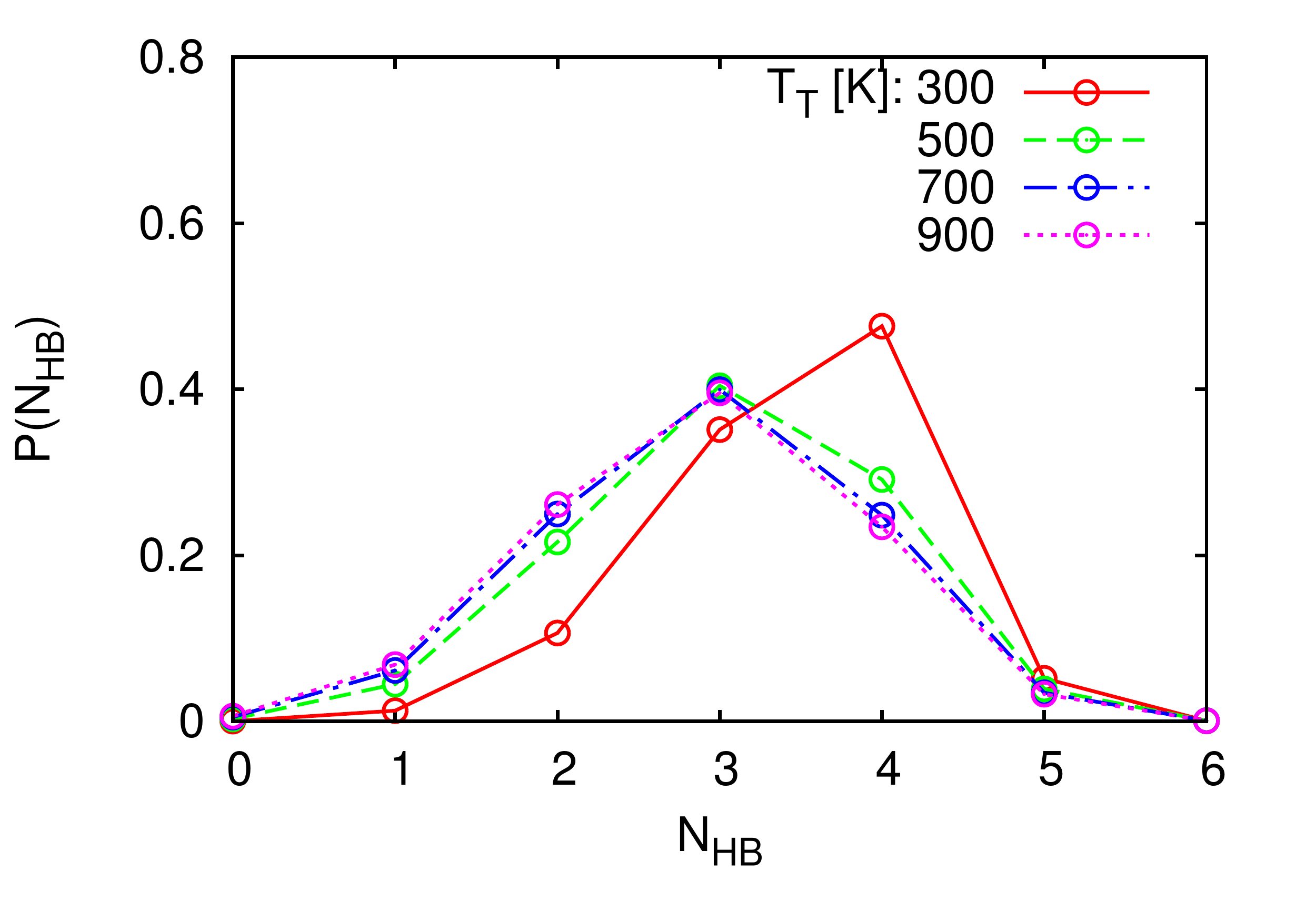}
\end{subfigure}
\caption{(Color online) Probability for a water molecule to form $N_\textrm{HB}$ hydrogen bonds, $P(N_\textrm{HB})$, at different $T_\textrm{R}$ (left-hand panel) and $T_\textrm{T}$ (right-hand panel). We count two water molecules as hydrogen-bonded if (i) the angle between the OH arm of one molecule and the O--O vector is less than $30^\circ$, and (ii) if the O--O distance is less than 3.3~{\AA} (position of the first minimum in $g_\textrm{OO}$). The color
code as for figure~\ref{pdf-1}.}
\label{fig-3}
\end{figure}

In figure~\ref{fig-3}, we show the probability $P(N_\textrm{HB})$ that
a water molecule forms $N_\textrm{HB}$ hydrogen bonds. Up to 500~K, the
probability distribution, $P(N_\textrm{HB})$, is practically the same for
the $T_\textrm{R}$ and $T_\textrm{T}$ dependence. Above $T_\textrm{R}$=500~K, the peak of this
distribution is gradually shifted toward smaller $N_\textrm{HB}$
values and there is an appreciable probability for only one hydrogen bond.
The situation is different for an increase of $T_\textrm{T}$
where the $P(N_\textrm{HB})$ distribution does not change much upon
an increase of the translational temperature above $T_\textrm{T}$=500~K.

All the evidence gathered so far suggests that hydrogen-bonding
is more sensitive to $T_\textrm{R}$, than to $T_\textrm{T}$ variations. The
reason for this has been discussed in reference
\cite{Bren2012} (page 024108-6). The formation of a hydrogen
bond requires two water molecules to meet at an appropriate
distance and a correct orientation. It is easy to understand that every
increase in $T_\textrm{R}$ (faster rotation) will make favorable orientations less
probable. At conditions of high $T_\textrm{R}$, the model molecules will
behave as independent rotators with the hydrogen bond
interaction energy approaching zero.

The mechanism causing two
molecules of water to lose the correlation upon the $T_\textrm{T}$
increase is different. In the latter case, the molecules should separate for a certain distance from each other, which is
not that easy because each molecule is caged by the others. By
increasing $T_\textrm{T}$, having in mind that we perform simulations at a constant
water density, we do not approach the ideal gas behaviour, even
at high temperatures ($T_\textrm{T}$=900~K) the system is under the effect
of strong repulsive interaction. An initial increase of $T_\textrm{T}$
(up to $T_\textrm{T}=500$~K) indicates that some hydrogen bonds are
broken and the radial order is disrupted to a certain extent.

\begin{figure}[htb!]
\centering
\begin{subfigure}[b]{0.45\textwidth}
	\centering
	\includegraphics[keepaspectratio=true,angle=270,width=\textwidth]{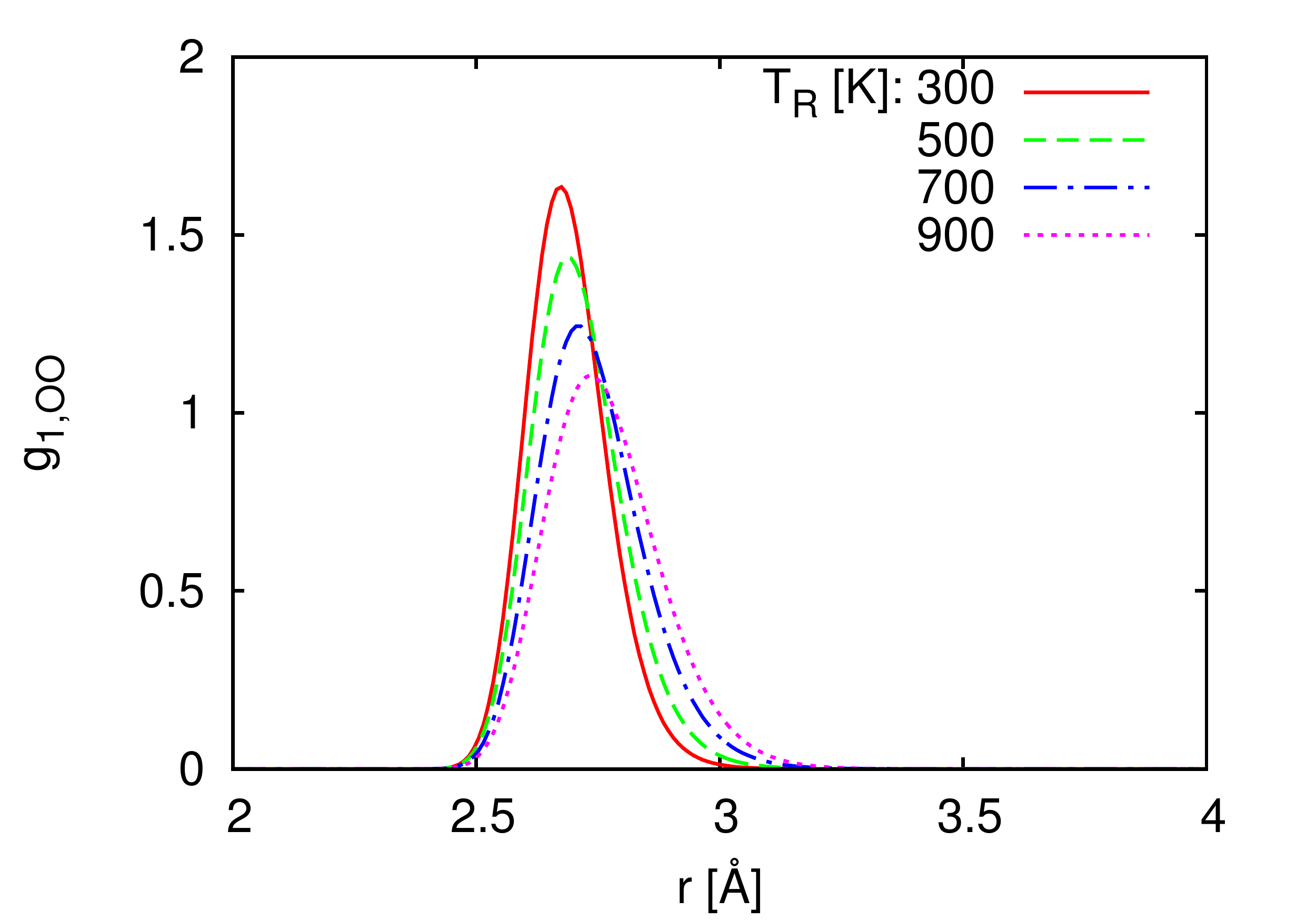}
\end{subfigure}%
\begin{subfigure}[b]{0.45\textwidth}
	\centering
	\includegraphics[keepaspectratio=true,angle=270,width=\textwidth]{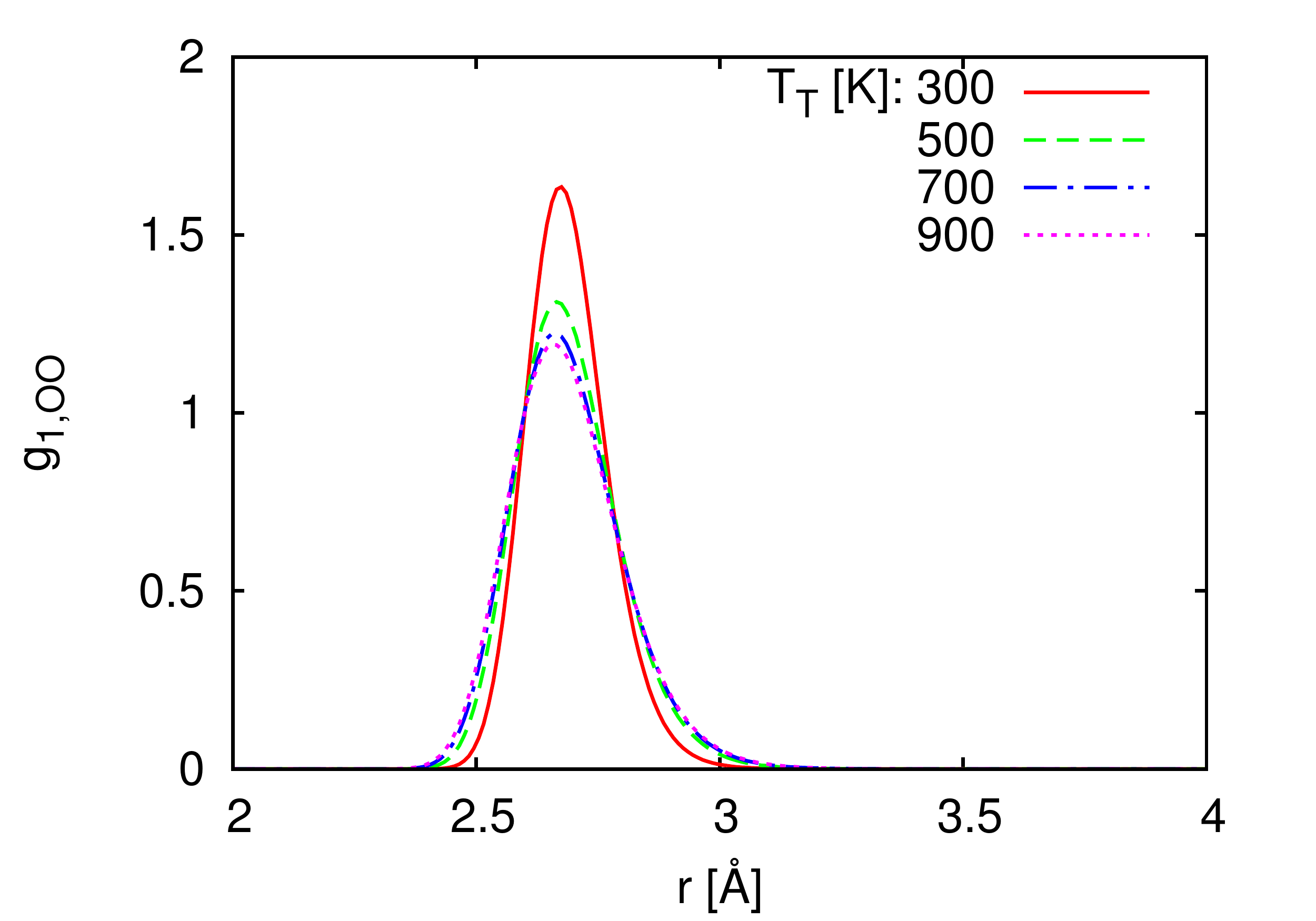}
\end{subfigure}
\begin{subfigure}[b]{0.45\textwidth}
	\centering
	\includegraphics[keepaspectratio=true,angle=270,width=\textwidth]{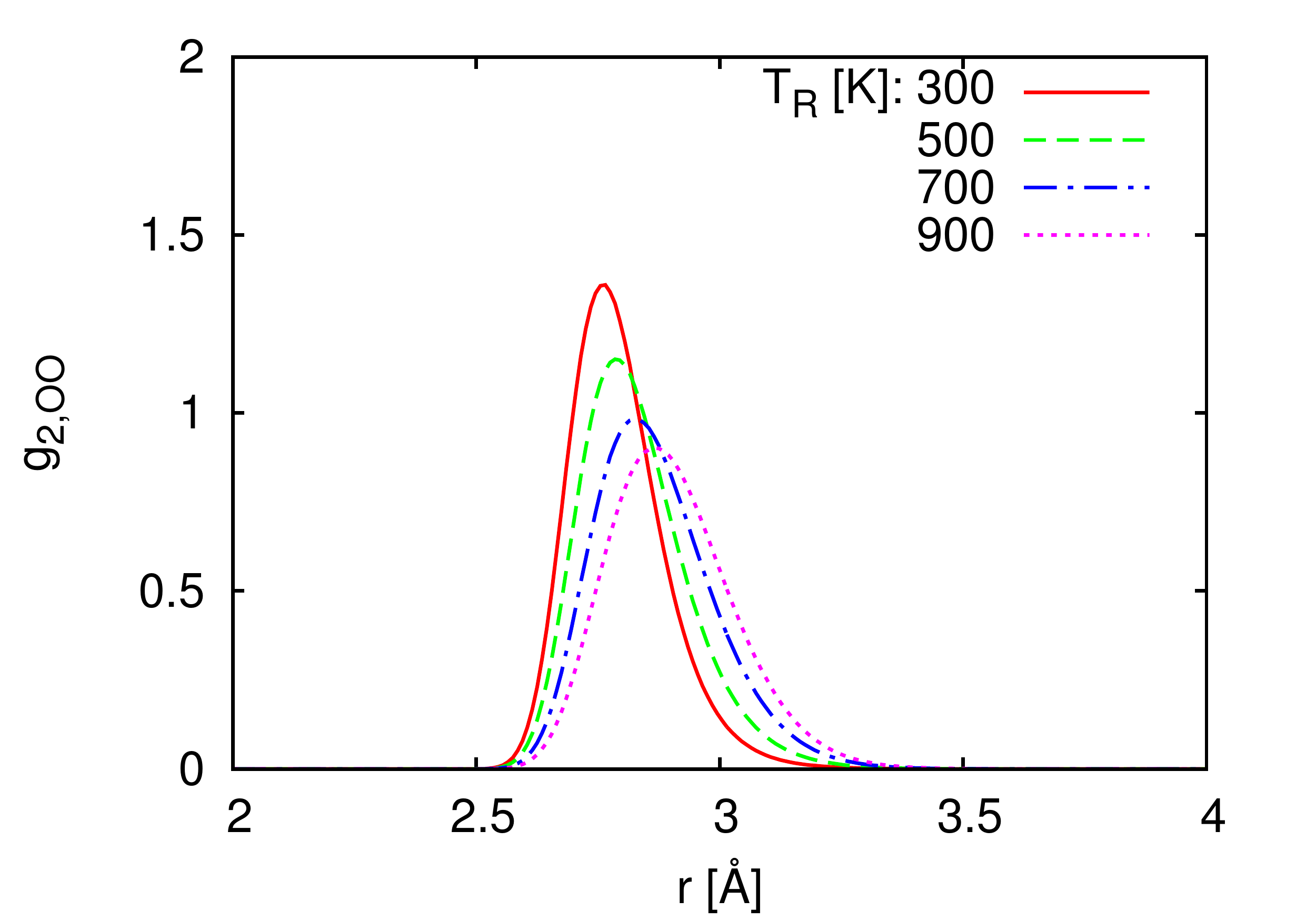}
\end{subfigure}%
\begin{subfigure}[b]{0.45\textwidth}
	\centering
	\includegraphics[keepaspectratio=true,angle=270,width=\textwidth]{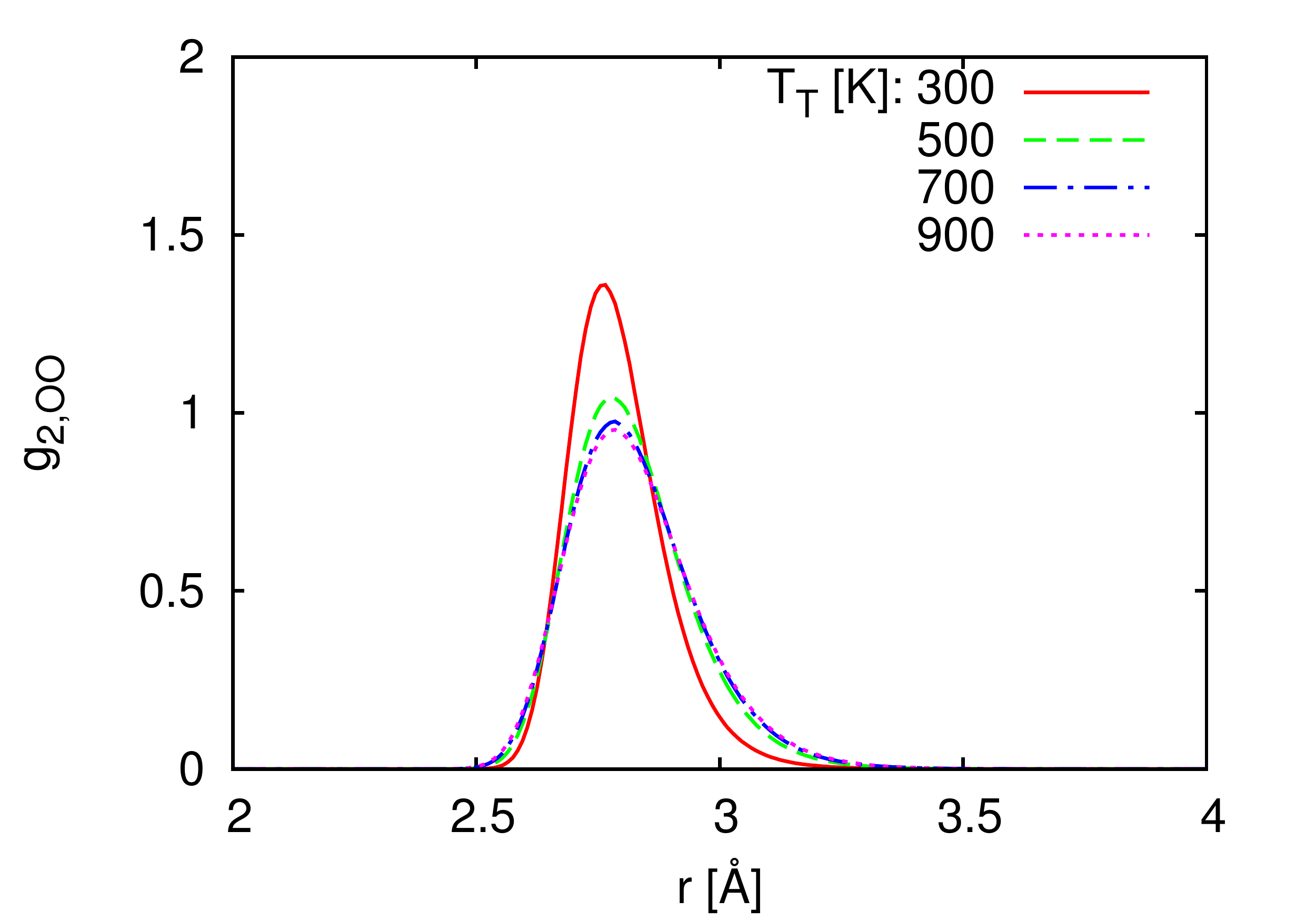}
\end{subfigure}
\begin{subfigure}[b]{0.45\textwidth}
	\centering
	\includegraphics[keepaspectratio=true,angle=270,width=\textwidth]{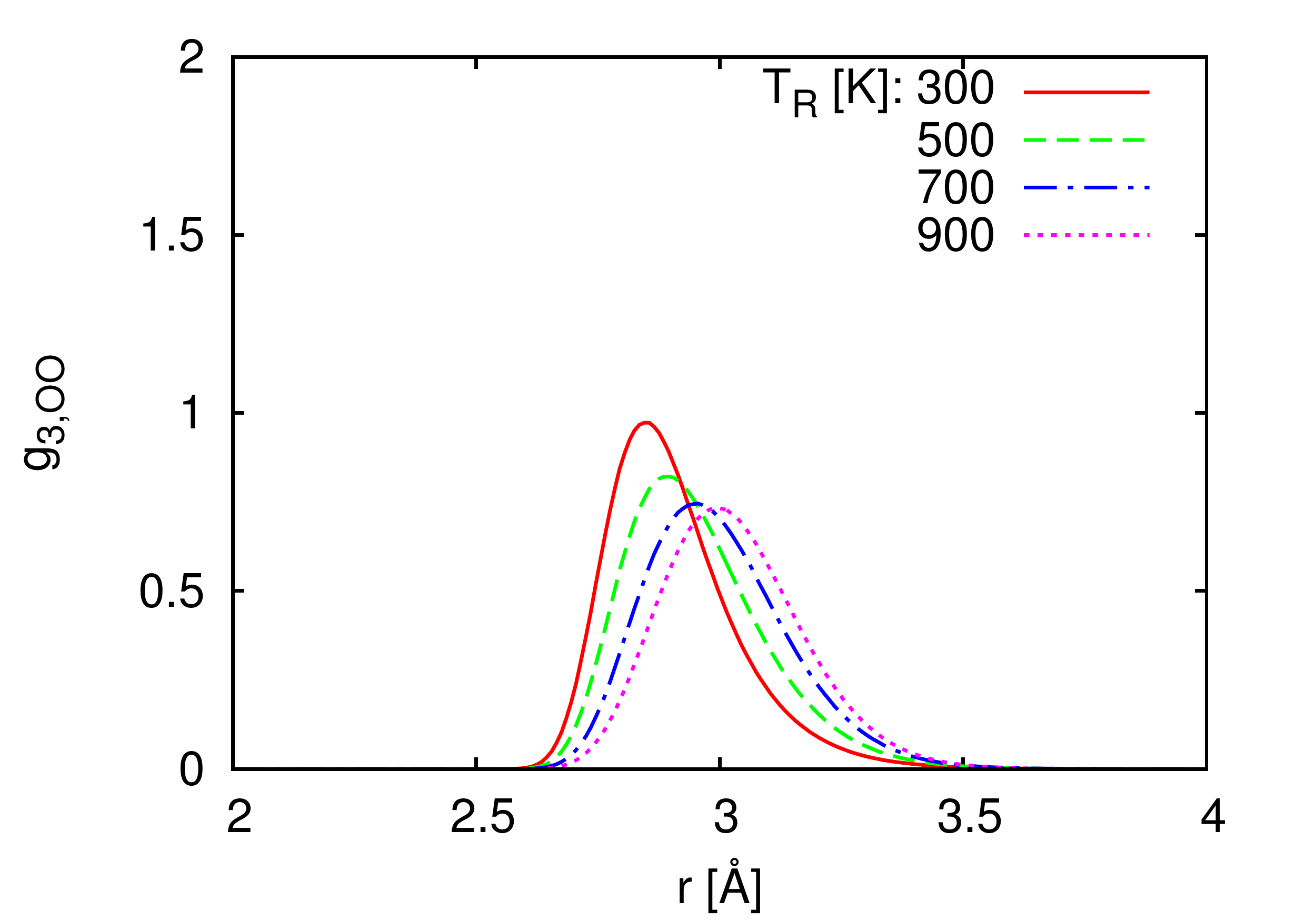}
\end{subfigure}%
\begin{subfigure}[b]{0.45\textwidth}
	\centering
	\includegraphics[keepaspectratio=true,angle=270,width=\textwidth]{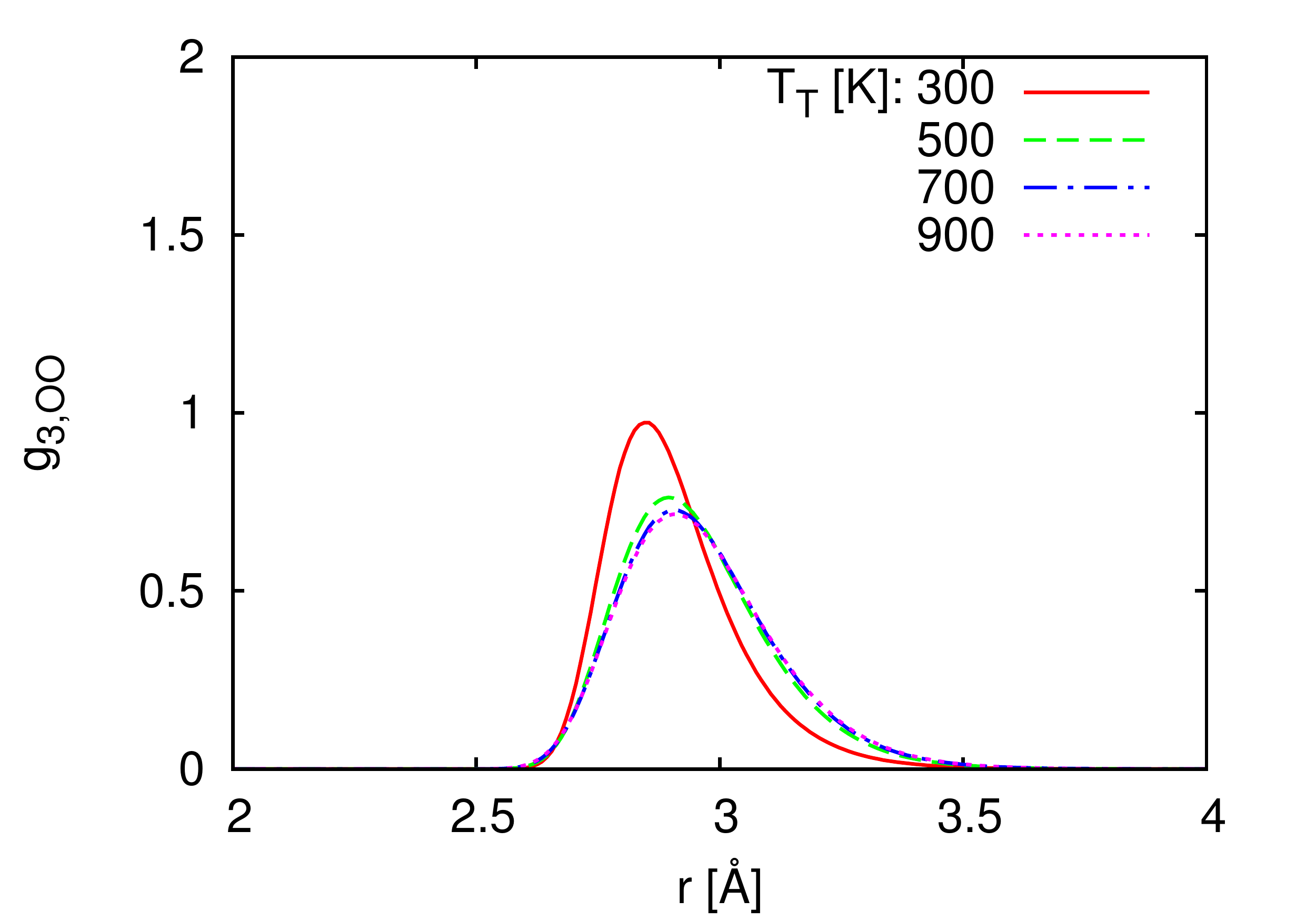}
\end{subfigure}
\begin{subfigure}[b]{0.45\textwidth}
	\centering
	\includegraphics[keepaspectratio=true,angle=270,width=\textwidth]{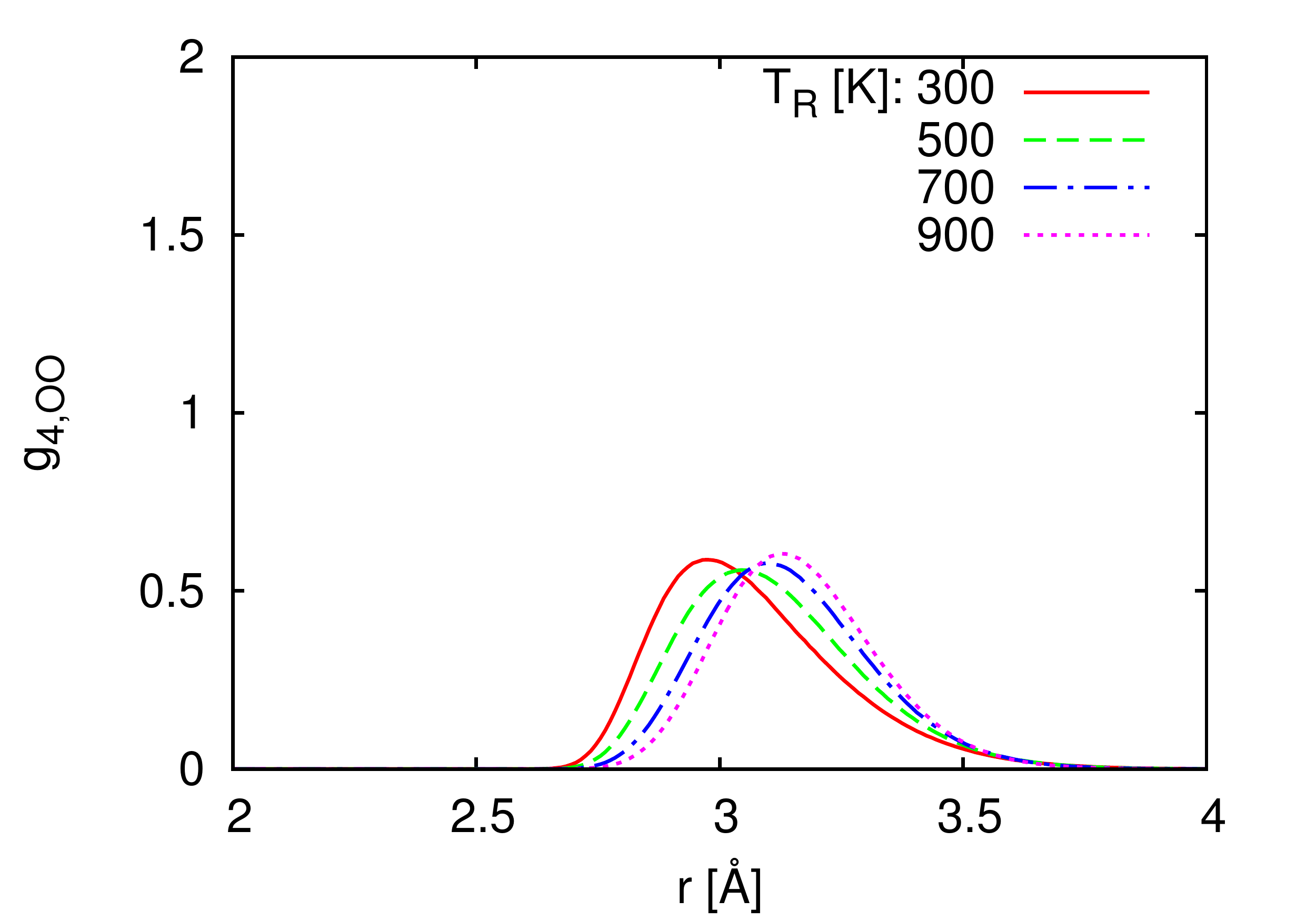}
\end{subfigure}%
\begin{subfigure}[b]{0.45\textwidth}
	\centering
	\includegraphics[keepaspectratio=true,angle=270,width=\textwidth]{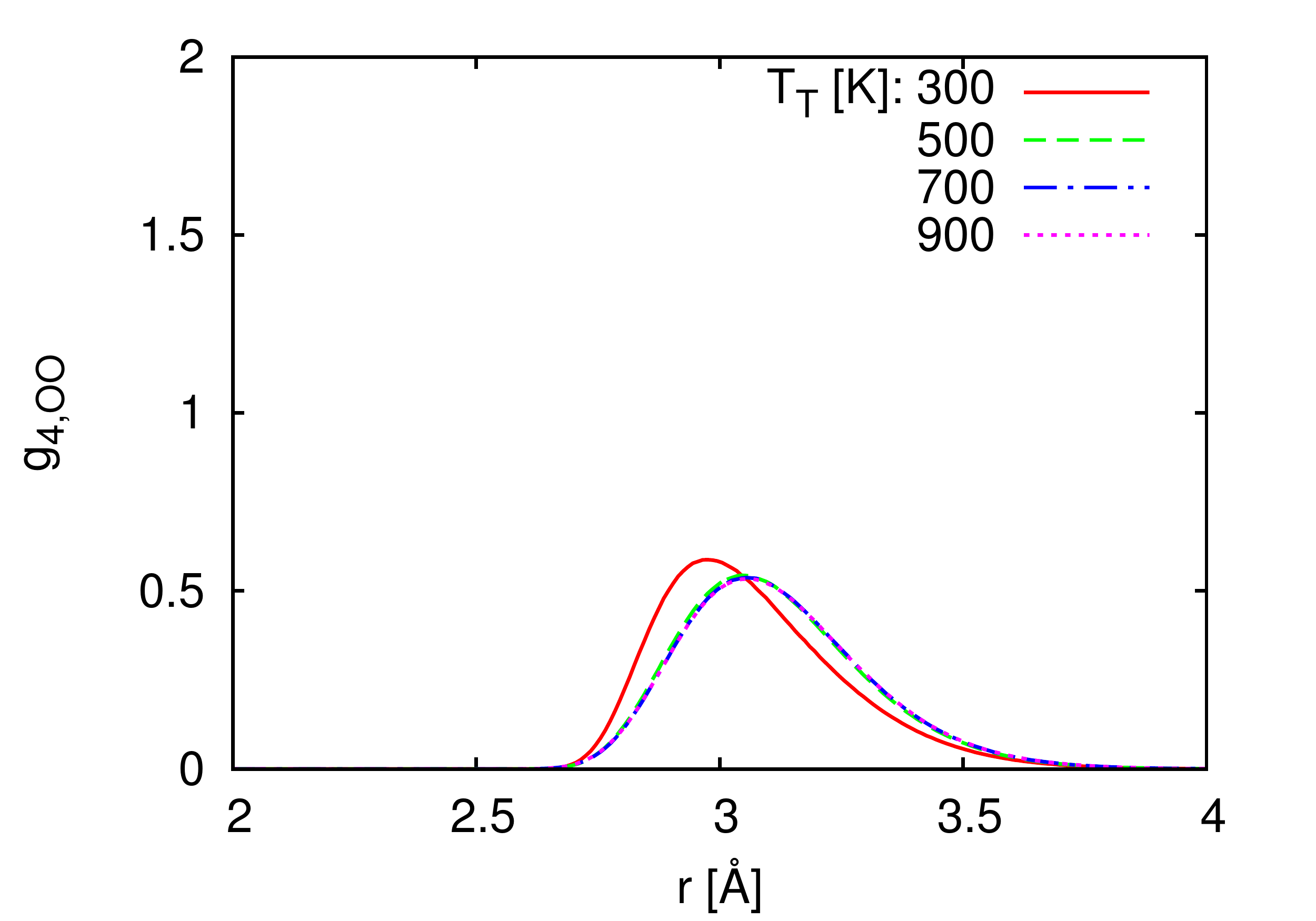}
\end{subfigure}
\caption{(Color online) Oxygen--oxygen radial distribution function for the first four
neighbours of the water molecule (from top to the bottom $g_1$ to $g_4$) at
different $T_\textrm{R}$ (left-hand panels) and $T_\textrm{T}$ (right-hand panels). The color code as for figure~\ref{pdf-1}.}
\label{neighbours}
\vspace{-4mm}
\end{figure}

A more detailed information on the water structure can be obtained
if the radial distribution function, $g_\textrm{OO}$, is divided into
the separate contributions of the neighboring molecules
coordinated on the chosen molecule. In our notation, $g_i$ is
the contribution of the $i$-th neighbor to the total pair
distribution function. These results are shown in figure~\ref{neighbours}, where the left-hand panels belong to different
rotational and the right-hand panels belong to different translational
temperatures. In case of $T_\textrm{R}$ increase, we observe the shifts
of the peaks toward larger distances. This applies to all four
nearest neighbors. At this density water molecules are closely
packed; while $\sigma_{WW}$ for this (SPC/E) model is 3.169~{\AA}, the position of the first peak is actually at around
$2.7$~{\AA}. Rising the $T_\textrm{R}$ causes for this distance to
increase, and the molecules distribute at larger average distances.
In contrast to this, an increase of the $T_\textrm{T}$ does not significantly change
the position of the first peak. This is yet another
consequence of the fact that faster rotation (increase in $T_\textrm{R}$)
disrupts the hydrogen bond network more efficiently than an increase
in $T_\textrm{T}$.

An interesting question, related to the changes of peak positions of various
$g_{i}$'s upon $T_\textrm{R}$ increase, has  been posed by an anonymous reviewer.
Figure~\ref{neighbours} indicates that the position of the first peak, $g_1$,
is less affected than the positions of more distant peaks,
such as $g_3$ and $g_4$. Why? Considering that $g_3$ measures the
contribution of the third neighbour to the oxygen--oxygen distribution function,
we may conclude that correlation between the central molecule and the third shell
molecules is significantly suppressed. In other words, the effect of the central
molecule does not propagate that far in the bulk as for $T_\textrm{R}=T_\textrm{T}=300$~K.
This is clearly shown in figure~\ref{pdf-1}; the structure of the solution as measured
by oxygen--oxygen pair distribution function, $g_\textrm{OO}$, becomes more
uniform upon an increase of either $T_\textrm{R}$ or $T_\textrm{T}$.  Furthermore, oscillations are
not just smaller at elevated temperature, they are also ``out of phase''.

\subsection{Probability of observing an empty cavity of a size of water}

\begin{figure}[!h]
\centering
\includegraphics[keepaspectratio=true,angle=270,scale=0.3]{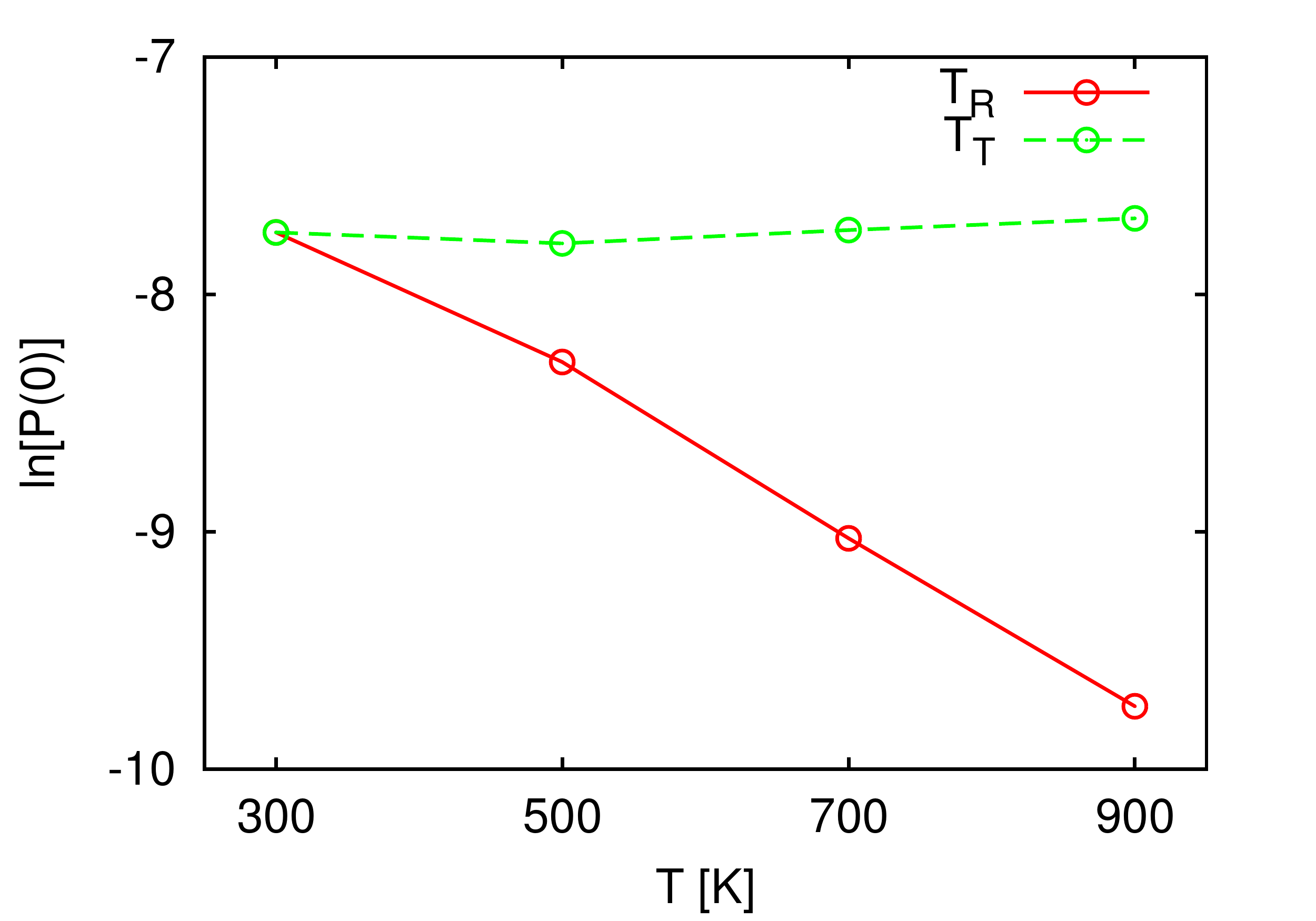}
\caption{(Color online) Logarithm of probability $P(0)$ of observing an empty spherical cavity of diameter 3~{\AA} in the  model water as a function of $T_\textrm{R}$ (solid, red) and $T_\textrm{T}$ (dashed, green).}
\label{prob0}
\end{figure}

The simulations presented in this paper are performed for a
constant number of particles in a given volume. An important
information on a liquid system is associated with the density
fluctuations.  In this subsection we studied the
probability distribution $P(N)$ for observing exactly $N$ water
molecules (actually their centers of gravity) in a spherical volume having
diameter 3~{\AA}. This probability distribution is approximately Gaussian \cite{Hummer1996} for small (molecular size)
volumes.
The width of the distribution reflects compressibility of the
liquid: a wider distribution means a somewhat higher
compressibility. Our simulations indicate that $P(N)$ curves change
only marginally upon an increase of $T_\textrm{R}$ and $T_\textrm{T}$. A notable exception is
the $P(0)$ value~--- i.e., the probability of observing an empty spherical cavity.
Accordingly, only the dependencies of $P(0)$ on $T_\textrm{R}$ and $T_\textrm{T}$ are shown
in figure~\ref{prob0}.
An increase  of $T_\textrm{R}$ from 300 to 900~K causes  a
decrease for about one order of magnitude for $P(0)$ (note the natural logarithm plotted
on $y$-axis in figure~\ref{prob0}). The density
fluctuations are suppressed due to a decreased number of
hydrogen bonds. On the other hand, the variations of $T_\textrm{T}$ have
little effect on $P(0)$.

\begin{figure}[!t]
\centering
\includegraphics[keepaspectratio=true,angle=270,scale=0.3]{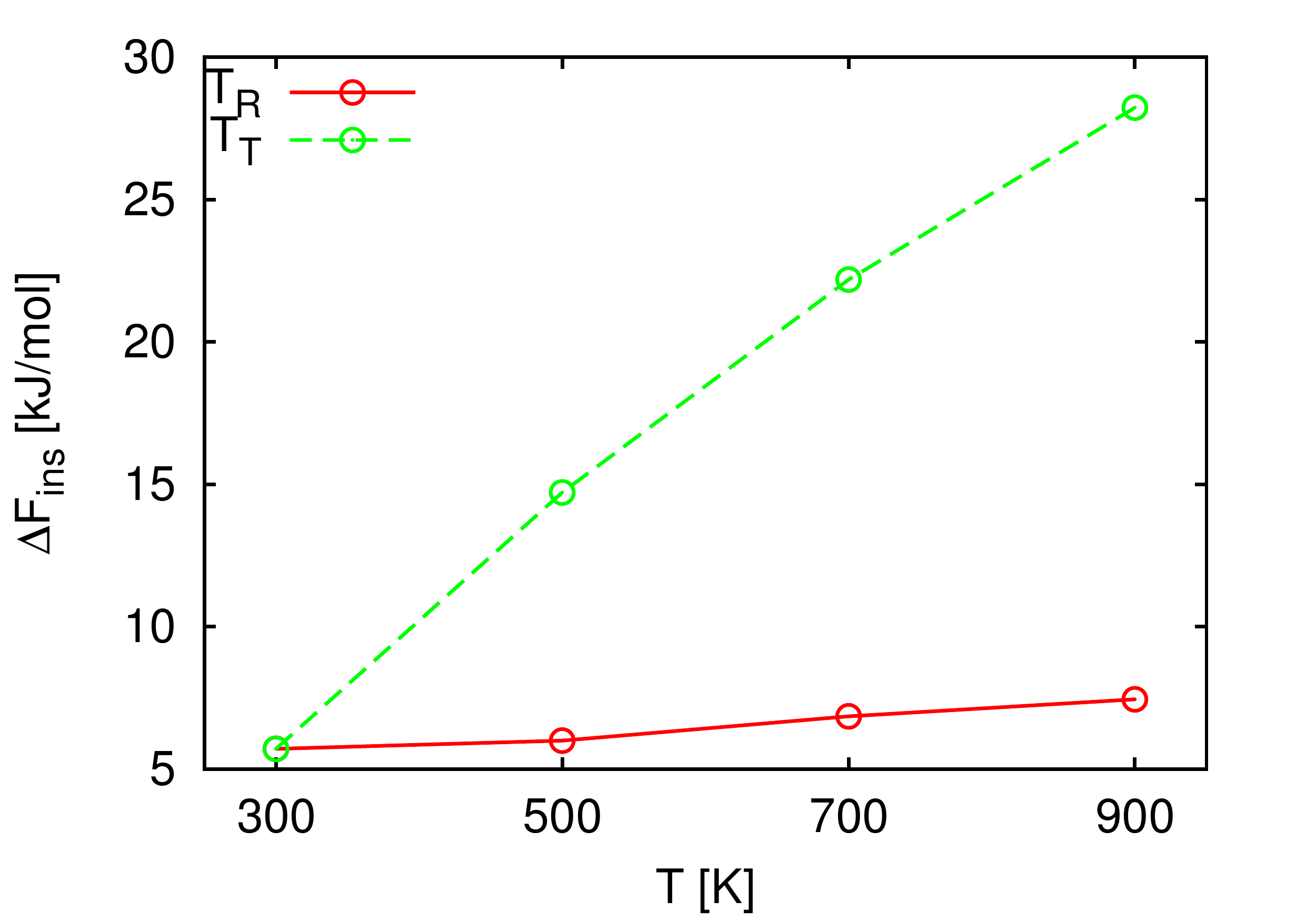}
\caption{(Color online) Free energy of the hydrophobe insertion as a function of $T_\textrm{R}$ (solid, red) or $T_\textrm{T}$ (dashed, green).}
\label{free}
\vspace{-2mm}
\end{figure}

Complementary to these calculations, we also performed the
thermodynamic integration to simulate the free energy of a
hydrophobe insertion, $\Delta F_\textrm{ins}$, into the model water (see figure~\ref{free}).
Hydrophobe was represented by the Lennard-Jones particle with
the well depth equal to that of the model water molecule and
$\sigma_{SS}=3$~{\AA}. Lorentz-Berthelot mixing rule was used
for cross interaction parameters. Under conditions where the hydrophobe--water
coupling is weak, we expect for the $\Delta
F_\textrm{ins}$ to have qualitatively the same features as the same
quantity for the hard-sphere solute.

While the $P(0)$ values decrease with an increase of $T_\textrm{R}$, the
insertion free energy is only slightly affected under such
conditions. On the other hand, $T_\textrm{T}$, which has virtually no
effect on the $P(N)$ distribution (see figure~\ref{prob0}), has
a relatively large effect on the $\Delta F_\textrm{ins}$. In both
cases, the free energy for the particle insertion, $\Delta
F_\textrm{ins}$, increases; it becomes increasingly more difficult
to add a hydrophobe to the system.

Here, we offer a qualitative explanation for this result. The insertion
free energy of the hard-sphere solute is related to the probability of
observing an empty spherical cavity, $P(0)$, in solvent \cite{Hummer1996}
(as usual $k_\textrm{B}$ is the Boltzmann constant):
\begin{equation}
\Delta F_\textrm{ins}=- k_\textrm{B}T \ln{P(0)}.
\label{eq_free}
\end{equation}
As the solute--water interaction depends only on the positions
of water molecules (and not on their orientations) and since
equation (\ref{eq_free}) applies to \emph{any} solvent (not
necessarily possessing rotational degrees of freedom), we might
expect that temperature in equation (\ref{eq_free}) refers to the
translational temperature. While the $P(0)$ values vary
substantially with $T_\textrm{R}$, the $ \ln P(0)$ only increases for
about 20~\% in magnitude, which is close to the relative
change of $\Delta F_\textrm{ins}$ with $T_\textrm{R}$. Much stronger increase
of $\Delta F_\textrm{ins}$ is observed upon the rise of $T_\textrm{T}$ and it may
be attributed to an increase of the pre-factor $k_\textrm{B}T$.

\subsection{Excess internal energy}

\begin{figure}[!b]
\centering
\includegraphics[keepaspectratio=true,angle=270,scale=0.3]{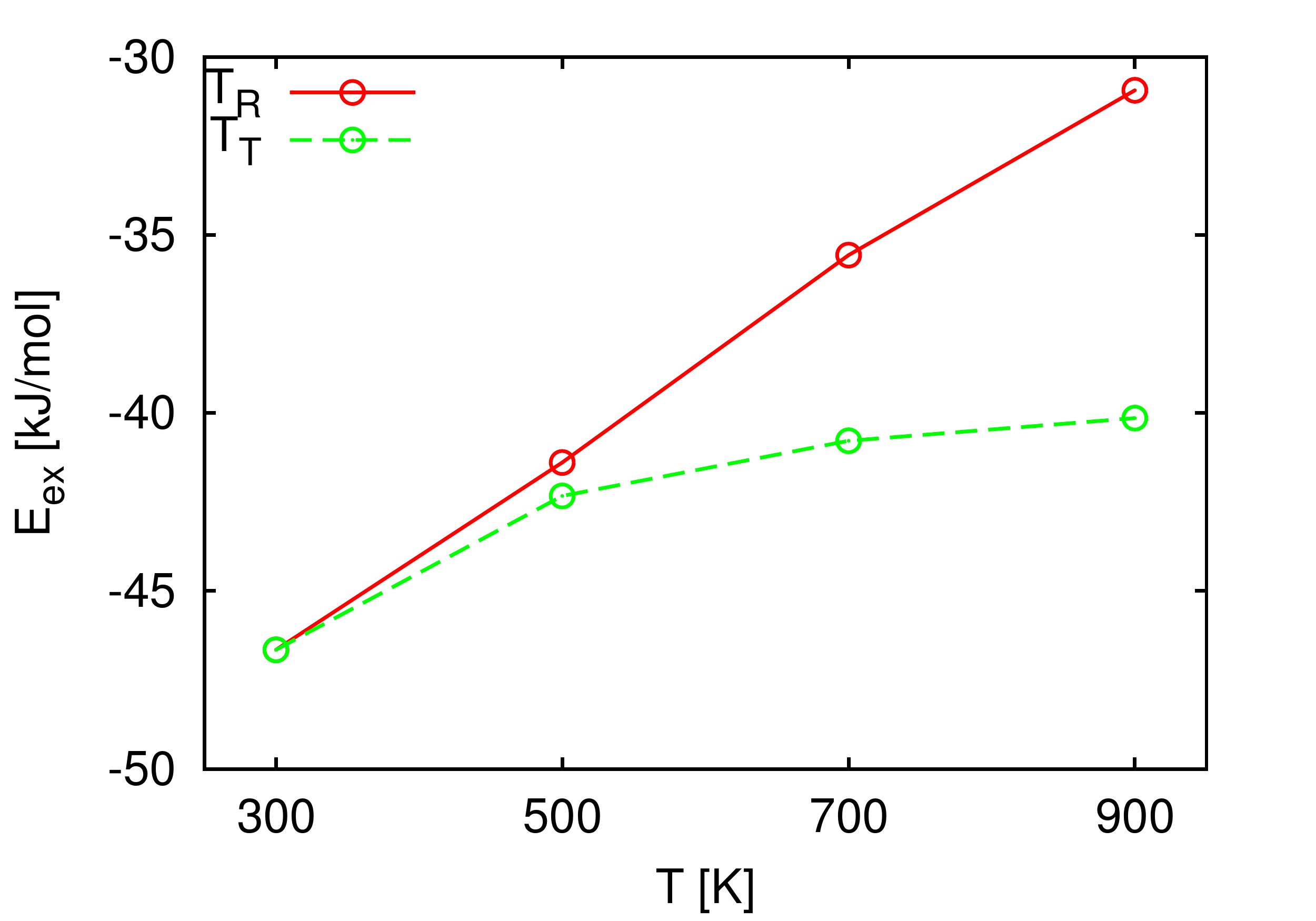}
\caption{(Color online) The excess internal energy per mol ($E_\textrm{ex}$) as a function of $T_\textrm{R}$ (solid, red) and $T_\textrm{T}$ (dashed, green).}
\label{excess}
\vspace{-2mm}
\end{figure}

It is of interest to examine the variations of the excess internal
energy ($E_\textrm{ex}$) per mol of water molecules as functions of $T_\textrm{R}$ or $T_\textrm{T}$ (figure~\ref{excess}).
We can see that elevation of $T_\textrm{R}$ causes an
almost linear increase of $E_\textrm{ex}$, while the same quantity changes less
and less if $T_\textrm{T}$ is increased.  The reason for such a behaviour lies, as
explained before, in different effects of the two degrees of freedom to hydrogen bonding.
These graphs also provide some insights into the $T_\textrm{R}$ and $T_\textrm{T}$
effects on the excess heat capacities (at constant $N$ and $V$),
$C_{v,r}^\textrm{ex}$ and $C_{v,t}^\textrm{ex}$. As we see, the rotational
contribution (derivative of the upper curve) is temperature
independent, while for the translational contribution
(derivative of the lower curve) the excess heat capacity goes
to zero for high $T_\textrm{T}$ values.

\subsection{Dynamics of model water molecule upon an increase of $T_\textrm{R}$ and $T_\textrm{T}$}

\begin{figure}[!t]
\centering
\includegraphics[keepaspectratio=true,angle=270,scale=0.3]{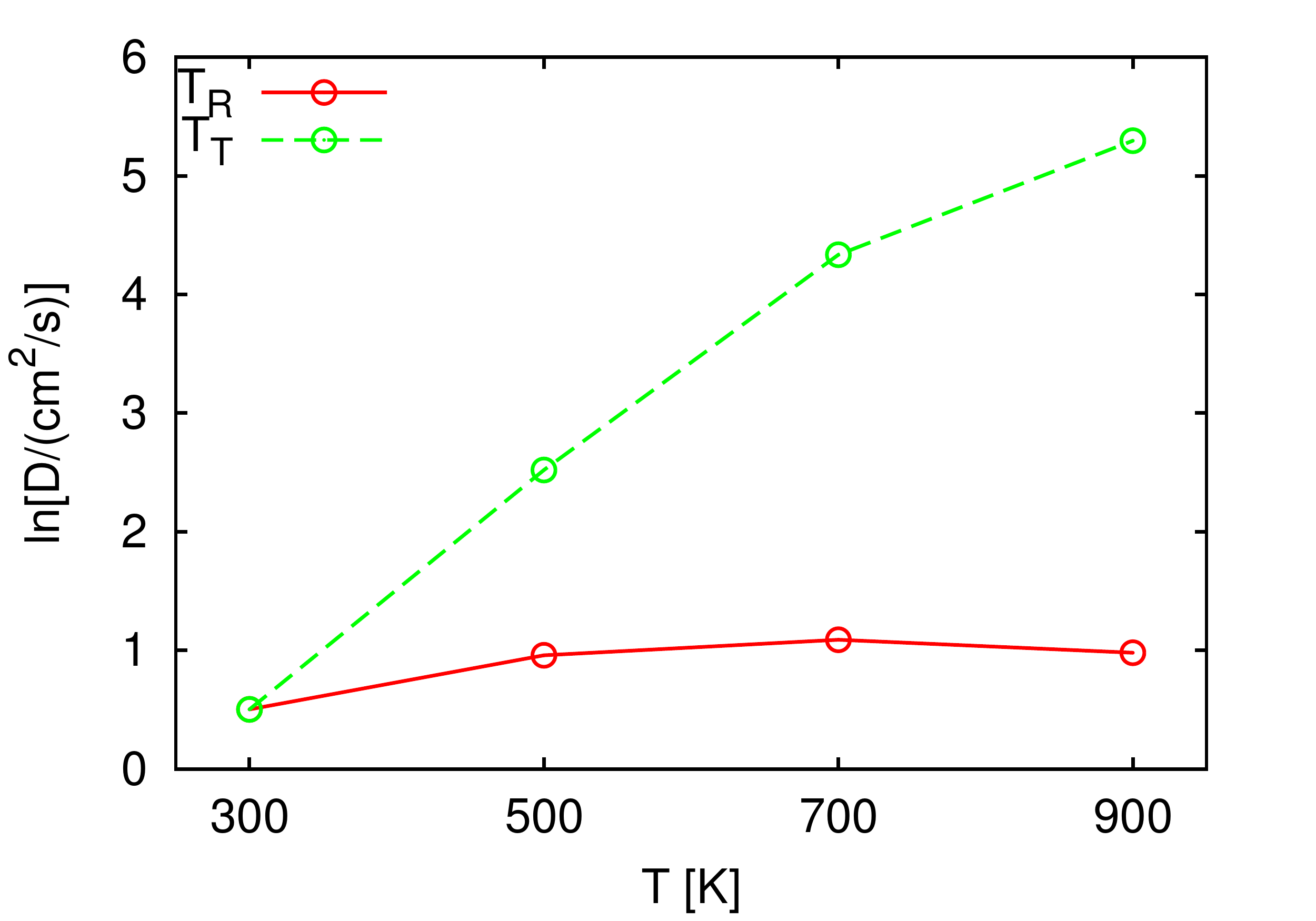}
\caption{(Color online) Diffusion coefficient as a function of $T_\textrm{R}$ (solid, red) or $T_\textrm{T}$
(dashed, green). Note the logarithmic scale on $y$-axis.}
\label{diffusion}
\end{figure}

How the different degrees of freedom affect the dynamics of
water molecules? We can see the answer  in figure~\ref{diffusion},
showing the effect of $T_\textrm{R}$ and $T_\textrm{T}$ on the (self) diffusion
coefficient of a model water molecule.

At first, an increase of $T_\textrm{R}$ causes a faster diffusion, which
can be attributed to the breaking of hydrogen bonds. As the
water molecules are more weakly bound, they tend to diffuse
faster. However, an increase of $T_\textrm{R}$ also means a decrease of
$P(0)$. Therefore, a random walk should proceed in smaller
steps, causing the diffusion to slow down. This effect seems to
dominate at the highest $T_\textrm{R}$ studied here. On the other hand,
increasing $T_\textrm{T}$, as expected, causes a monotonous increase of the
diffusion coefficient in this temperature interval.

\section{Conclusions}
\label{sec:conclusion}

It is well known that microwaves induce heating by excitation
of rotational motion in water molecules and that the rotational
kinetic energy is then transferred to the translational degrees
of freedom.  The microwave irradiation is important in biology:
there are reports that microwaves enhance the
folding and unfolding kinetics of globular proteins
\cite{Bohr2000,Bohr2000a,Chinnadayyala}, as well as protein
aggregation \cite{Pomerai2003}.

The effect of microwaves on the properties of liquids and
solutions can be studied by non-equilibrium molecular
dynamics simulations \cite{Tanaka2007}. In such studies, the
electric field is modelled explicitly. Following the ideas,
proposed in references \cite{Bren2010,Bren2012}, we perform
calculations in which the rotational degree of freedom has
the temperature different from the translational one. In a previous
paper \cite{Mohoric2014}, we examined the effect of such non-equilibrium
conditions on the solvation of simple solutes.
In the present contribution, we studied the properties of a model water under
non-equilibrium conditions where the translational and
rotational temperatures vary independently. The main
conclusions of this work are: (i) an increase of the
rotational temperature causes the first peak of $g_\textrm{OO}$ to
move toward larger distances, while its position remains
unchanged upon $T_\textrm{T}$ increase. (ii) The height of the first
peak in the hydrogen--oxygen distribution function is much more
sensitive on the $T_\textrm{R}$ than on the $T_\textrm{T}$ variations. (iii) At
high rotational temperatures, the number of hydrogen bonds may
drop as low as to one. (iv) Free energy of the hydrophobe
insertion does only marginally depend on the rotational
temperature. (v) Water diffusion coefficient measured via the
mean square displacement is, in contrast to the $T_\textrm{T}$
dependence, a non-monotonous function of the rotational
temperature. These and other effects observed upon varying the rotational
temperature may significantly affect the properties of water as
solvent, which has to a certain degree been examined in reference
\cite{Mohoric2014}. The work exploring such effects on polymer
and polyelectrolyte conformations in aqueous solution is currently
underway~--- preliminary results clearly point toward a high
probability of polymer collapse in cases of rotational heating.

In the end, we shall conclude that, despite important
developments following reference \cite{Barker}, there is still
ample room for improvements in theory and simulation of liquids, of which,
water is by far most important.

\section*{Acknowledgements}
This study was supported by the Slovenian Research Agency fund
(ARRS) through the Program 0103--0201, Project J1--4148,
GM063592 grant of NIH U.S.A., and the Young Researchers Program
(T. M.) of Republic of Slovenia.

\ukrainianpart

\title{Вплив трансляційних та обертових ступенів вільності на властивості моделі води}

 \author{T. Мохоріч, Б. Грібар-Лі, В. Влахі}
 \address{Факультет хімії і хімічної технології, Унiверситет Любляни, Ашкерчева
 5, SI–1000 Любляна, Словенія}

\makeukrtitle

\begin{abstract}
 Для вивчення впливу трансляційних та обертових ступенів вільності на структуру води з фіксованою густиною
 використано метод молекулярної динаміки з двома окремими термостатами для обертових та трансляційних рухів.
 Для опису молекул води було використано модель SPC/E. Результати показують, що збільшення обертової температури,
 $T_\textrm{R}$, призводить до значного розриву водневих зв'язків. Цілком зворотнім є випадок, принаймні не до такої міри, коли збільшується
 трансляційна температура, $T_\textrm{T}$. Ймовірність знайти сферичну порожнину (відсутня молекула води)
 заданого розміру значно зменшується при збільшенні $T_\textrm{R}$, але це має лише незначний вплив на вільну енергію гідрофобної вставки.
 Надлишкова внутрішня енергія збільшується пропорційно до збільшення $T_\textrm{R}$, в той час, як $T_\textrm{T}$ дає значно менший
 ефект при високих температурах. Коефіцієнт дифузії води має немонотонну поведінку при збільшенні обертової температури.

\keywords структура води, ступені вільності, молекулярна динаміка

\end{abstract}

  \end{document}